\documentstyle[amstex]{mn}

\input epsf
\def\plotone#1{\centering \leavevmode
\epsfxsize=\columnwidth \epsfbox{#1}}
\def\plottwo#1#2{\centering \leavevmode
\epsfxsize=.99\columnwidth \epsfbox{#1} \hfil
\epsfxsize=.99\columnwidth \epsfbox{#2}}
\def\plotonew#1{\centering \leavevmode
\epsfxsize=1.99\columnwidth \epsfbox{#1}}
\def\plotthree#1#2#3{\centering \leavevmode
\epsfxsize=.99\columnwidth \epsfbox{#1}\hfil
\epsfxsize=.99\columnwidth \epsfbox{#2} \hfil
\epsfxsize=.99\columnwidth \epsfbox{#3}}

\title{Buoyant radio plasma in clusters of galaxies} 
\author[]
       {M. Br\"uggen$^{1,2}$, C.R. Kaiser$^{1,3}$  \\
        $^1$ Max-Planck-Institut f\"ur Astrophysik,
Karl-Schwarzschild-Str.1, 85740 Garching, Germany\\ 
$^2$ Churchill College, Storey's Way, Cambridge, CB3 0DS, UK\\
$^3$ current address: Department of Physics \& Astronomy, University of Southampton,
University Road, Southampton SO17 1BJ, UK}

\pagerange{\pageref{firstpage}--\pageref{lastpage}}
\pubyear{2000}

\begin{document}

\maketitle

\begin{abstract}
Radio galaxies are known to inflate lobes of hot relativistic plasmas
into the intergalactic medium. Here we present hydrodynamical and
magnetohydrodynamical simulations of these hot plasma bubbles in FRII
objects. We focus on the later stages of their evolution after the jet
has died down and after the bow shock that surrounded the lobes at
earlier stages has vanished. We investigate the evolution of the
plasma bubbles as they become subject to Rayleigh-Taylor
instabilities. From our simulations we calculate the radio and X-ray
emissivities of the bubbles and discuss their appearance in
observations. Finally, we investigate the influence of large-scale
magnetic fields on the evolution of the bubbles. The issues of
re-acceleration and diffusion of relativistic particles are briefly
discussed.
 
\end{abstract}


\section{Introduction}

The jets of radio galaxies and radio-loud quasars are known to inflate
large lobes of radio plasma in the Inter Galactic Medium (IGM). These
regions appear either as well-defined cavities in the case of Fanaroff -
Riley (FR) type II objects (Fanaroff \& Riley 1974)\nocite{fr74} or
form broad, more diffuse outflows in FRI objects. In this paper we
concentrate on the evolution of the radio lobes of the more powerful
FRII objects. In our view, the dynamics of FRIs are more complex and we
will refer a discussion of these to a future paper.\\

The standard model for FRIIs explains the lobes as overpressured
bubbles that expand supersonically into the surrounding intergalactic
medium (IGM) (Scheuer 1974)\nocite{ps74}. In principle, the lobes are
unstable because they are surrounded by the denser, unperturbed IGM in
the gravitational potential well of the host galaxy. This situation
should cause the buoyant lobes to rise through the ambient medium
(e.g. Gull \& Northover 1973\nocite{gn73}). However, as long as the
bow shock, caused by their supersonic expansion, protects the radio
lobes, buoyant forces are not important for the dynamics of the lobes
(Scheuer 1974\nocite{ps74}). Analyses of spectral ageing show that the
jet phase may last for a few $10^8$ years (e.g. Alexander \& Leahy
1987\nocite{al87}) after which the jets presumably stop supplying
energy to the lobes. After that, they may still be overpressured with
respect to the IGM and continue to expand, driving a bow shock. This
phase resembles that of a strong explosion or supernova remnant
(e.g. Sedov 1959)\nocite{ls59}.\\

The continued expansion of the lobes will bring them into pressure
equilibrium with the IGM. At this point, the bow shock vanishes and
the lobes become susceptible to Rayleigh-Taylor instabilities. Because
of the absence of strong surface tension in the radio lobes, this
situation is reminiscent of the later phases of strong (nuclear)
explosions in the Earth's atmosphere. In both cases, the hot rarefied
gas forms a torus which, in its wake, drags material up into the
atmosphere. Thus the bubble develops a shape that resembles a
mushroom. This was discussed in connection with the large scale radio
structure of M87 by B{\"o}hringer et al. (1995)\nocite{bnbf95},
Churazov et al. (2000a)\nocite{cfjb00} and Churazov et
al. (2000b)\nocite{cbkbf00}.\\

In this paper we extend the hydrodynamical simulations for the
specific case of M87 presented in Churazov et al. (2000b)
\nocite{cbkbf00} to more general initial configurations. Moreover, we
performed MHD simulations to study the influence of a magnetic field
that is initially aligned with the physical boundary of the radio
lobes. We also present predictions for the radio synchrotron emission
from the magnetised plasma within the lobes. Finally, we demonstrate
that neither re-acceleration processes nor inhomogenous magnetic
fields in the radio lobes significantly alter the emission
properties of the lobes unless the diffusion of particles from weak
fields to high field regions is slow. \\

The paper is organised as follows: We discuss our methods and
assumptions in Sec. 2. The results of the simulations are presented in
Sec. 3, their observational consequences in Sec. 4 and the conclusions
in Sec. 5.

\section{Method}

\subsection{Magneto-hydrodynamical simulations}
\label{sec:simu}

The simulations were obtained using the ZEUS-3D code which was
developed especially for problems in astrophysical hydrodynamics
(Stone \& Norman 1992a, b).  The code uses finite differencing on a
Eulerian or pseudo-Lagrangian grid and is fully explicit in time. It
is based on an operator-split scheme with piecewise linear functions
for the fundamental variables. The fluid is advected through a mesh
using the upwind, monotonic interpolation scheme of van Leer.  The
magnetic field is evolved using a modified constrained transport
technique which ensures that the field remains divergence-free to
machine precision. The electromotive forces are computed via upwind
differencing along Alfv\'en characteristics. For a detailed
description of the algorithms and their numerical implementation see
Stone \& Norman (1992a, b).

As initial model we employed an isothermal model, where the density
is given by the $\beta$-model, i.e.

\begin{equation}
\rho(r)=\frac{\rho_0}{[1+(r/a_0)^2]^{3\beta /2}} .
\end{equation} 
We assumed $\beta =0.5$, $\rho_0 = 1.6\ 10^{-27}$ g cm$^{-3}$ and a
core radius of $a_0=200$ kpc, which are typical values for clusters of
galaxies (e.g. Jones \& Forman 1984\nocite{jf84}).  From this, the
gravitational potential was calculated such as to keep the background
density in hydrostatic equilibrium. The gravitational potential
remained fixed during the simulation, which saved us solving Poisson's
equation as part of the simulation.  We chose dimensionless units,
which are defined by $a_0=1$, $\rho_0 =1$ and a gravitational constant
of $G=1$. In our simulations we employed an ideal gas equation of
state and we ignored the effects of rotation and radiative processes.

The simulations were performed in 2D on a spherical grid
($r$,$\theta$) with 200 $\times$ 200 grid points. The grid span 1 Mpc
in the radial direction (400 kpc for model D - see below) and $\pi$
rad in the angular direction.

At the centre, a hot bubble of various geometrical shapes (see below)
was set up. It was made buoyant by lowering its density with respect
to the background density by a factor of 100 and simultaneously
raising the temperature by the same factor. Thus the bubble is in
pressure equilibrium with its surroundings.  The gas is treated as
a single fluid and is assumed to obey a polytropic equation of state
with $\gamma =5/3$. The bubble was filled with 3000 tracer particles
that are advected with the fluid. \\

The low frequency radio lobes of active FRII-type radio sources have
elliptical shapes similar to cigars (e.g. Blundell, Kassim \&
Perley 2000\nocite{bkp00}). We also chose elliptical shapes for
the initial configurations of our buoyant bubbles.  Moreover, it is
observed that some radio lobes are ``pinched'': The weakening of
the bow shock towards the end of the overpressured phase is not an
instantaneous process. The sideways expansion of the lobes proceeds
more slowly than the expansion in the direction of the jet. Moreover,
the regions of the lobe closest to the jet are located in the regions of highest
pressure. Therefore, it is likely that the bow shock will start to
weaken towards the centre of the radio source leading to a pinched
appearance (e.g. Baldwin 1982\nocite{jb82}). We address the effect of
such a pre-pinch in one of our models.

In our purely hydrodynamical simulations, we assume that the magnetic
field is tangled on scales much smaller than the numerical resolution
and that, consequently, it only plays a passive role in our
simulations. It has been suggested that the magnetic field may be
compressed at the edges of the radio lobes of active FRIIs. This may
lead to a more ordered field configuration with the field lines
preferentially aligned along the lobe surface (Laing 1980\nocite{rl80}
and references therein). We used the MHD capabilities of ZEUS-3D to
investigate such configurations.

In total we simulated six different bubble configurations:

\begin{itemize}
\item {\em Model A:}\/ Spherical bubble with a radius of 200 kpc
equivalent to the core radius, $a_0$.
\item {\em Model B:}\/ Elliptical bubble with its major axis oriented
along the external density gradient. The major axis has a length of
400 kpc and the aspect ratio of the ellipse is 4.
\item {\em Model C:}\/ `Pinched' elliptical bubble, whose surface is described
by $r=60\ {\rm kpc}/(1.+0.8\cos(\theta+\pi))$.
\item {\em Model D:}\/ Small elliptical bubble with a major axis of
100 kpc and an aspect ratio of 4. Note
that the major axis of this bubble only extends to half the core radius $a_0$.
\item {\em Model Bm1:}\/ Like Model B but with a large-scale magnetic
field which is roughly aligned with the bubble surface and which is at
the same time divergence-free. The field configuration is shown in
Fig. \ref{fig:modBm1} Energy density of the magnetic field
corresponds to about 10\% of the gas pressure. 
\item {\em Model Bm2:}\/ Like Model Bm1 but with a stronger magnetic field
of energy density equivalent to 30\% of the gas pressure.
\end{itemize}

\subsection{Radio emission from the simulated bubbles}
\label{sec:radsim}

At the start of our simulations, the radio lobes are at least partly
filled with relativistic, magnetised plasma. This material will emit
synchrotron radiation. The simulations assume a single,
non-relativistic fluid throughout the computational grid. In order to
approximate the evolution of the relativistic plasma, we need to make
the following assumptions (see also Churazov et
al. 2000b\nocite{cbkbf00}):

\begin{itemize}
\item In Models A through D we assume that the magnetic field within the
rarefied bubbles is tangled on small scales. This allows us to
treat the field together with the relativistic particles as a
`relativistic fluid' with an adiabatic index of $\Gamma = 4/3$. The
relativistic fluid is confined to small patches intermixed with the
thermal fluid. This implies that the two components of the bubbles are
mixed macroscopically but not on microscopic scales. In other words,
the magnetic field can provide a kind of surface tension on its
tangling scale.

\item The patches of relativistic fluid are in pressure equilibrium
with the thermal fluid. Thus they expand and contract adiabatically
but do not influence the overall dynamics of the fluid flow. 

\item The relativistic plasma and the magnetic field are contained in
the rarefied, buoyant bubble. The thermal pressure acting on the
relativistic plasma is balanced by the sum of the energy density of
the tangled magnetic field, $u_{\rm B}$, and that of the relativistic
particles, $u_{\rm e}$. Initially, we assume a power law for the energy
distribution of the relativistic particles with a high-energy
cut-off at $\gamma _{\rm max}$. In terms of the particle Lorentz
factor, this can be written as $n_{\rm e} d\gamma = n_{\rm o} \gamma
^{-p} d\gamma$, where we assumed $p=2$ and $\gamma _{\rm max}
=10^6$. At the start of the simulations we choose a value for the
energy density of the magnetic field, $u_{\rm B}$, which is uniform
throughout the bubble. The normalisation constant, $n_{\rm o}$, is
then fixed by the requirement of pressure equilibrium with the thermal
fluid.\\

\end{itemize}

For models Bm1 and Bm2 the magnetic field is evolved explicitly by the
MHD module of the ZEUS code. In these models we make the same
assumptions as in the other models except those for the magnetic
field.\\

Adiabatic expansion, synchrotron radiation and inverse Compton
scattering of the Cosmic Microwave Background (CMB) radiation lead to
changes of the energy distribution of the relativistic particles. The
first two processes depend on the local conditions of the relativistic
plasma. Furthermore, all energy losses are cumulative. In order to
determine the energy distribution of the relativistic particles at a
particular time at a given point in the simulation, we need to know
its entire `loss history'. This is achieved by tracking the conditions
at the position of the tracer particles which are advected with the
fluid flow. Assuming pressure equilibrium and adiabatic behaviour of
the relativistic fluid, we can thus calculate the energy density of
the magnetic field, $u_{\rm B} (t) = u_{\rm B} (t_{\rm o}) [p_{\rm th}
(t) / p_{\rm th} (t_{\rm o})]$, and the normalisation of the particle
energy spectrum, $n_{\rm o} (t) = n_{\rm o} (t_{\rm o}) [p_{\rm th}
(t) / p_{\rm th} (t_{\rm o})]^{3/4}$. The latter is only correct if
the diffusion of relativistic particles can be neglected. Using this
equation and interpolating linearly between computational time
steps, the energy distribution of the relativistic particles can be
constructed iteratively (e.g. Kaiser, Dennett-Thorpe \& Alexander
1997\nocite{kda97a}). Together with the local energy density of the
magnetic field, it is straightforward to calculate the synchrotron
emissivity at any given position of a tracer particle (e.g. Longair
1991\nocite{ml91}). In all simulations with elliptical bubbles the
initial symmetry axis along the major axis of the ellipse is preserved
at later times. Assuming rotational symmetry about this axis, we can
obtain a 3-dimensional model of the synchrotron emissivity of the
flow. Surface brightness maps are computed by line of sight
integration assuming optical thin conditions.

\section{Results of the simulations}

\subsection{The role of asymmetry}

\begin{figure}
\plotone{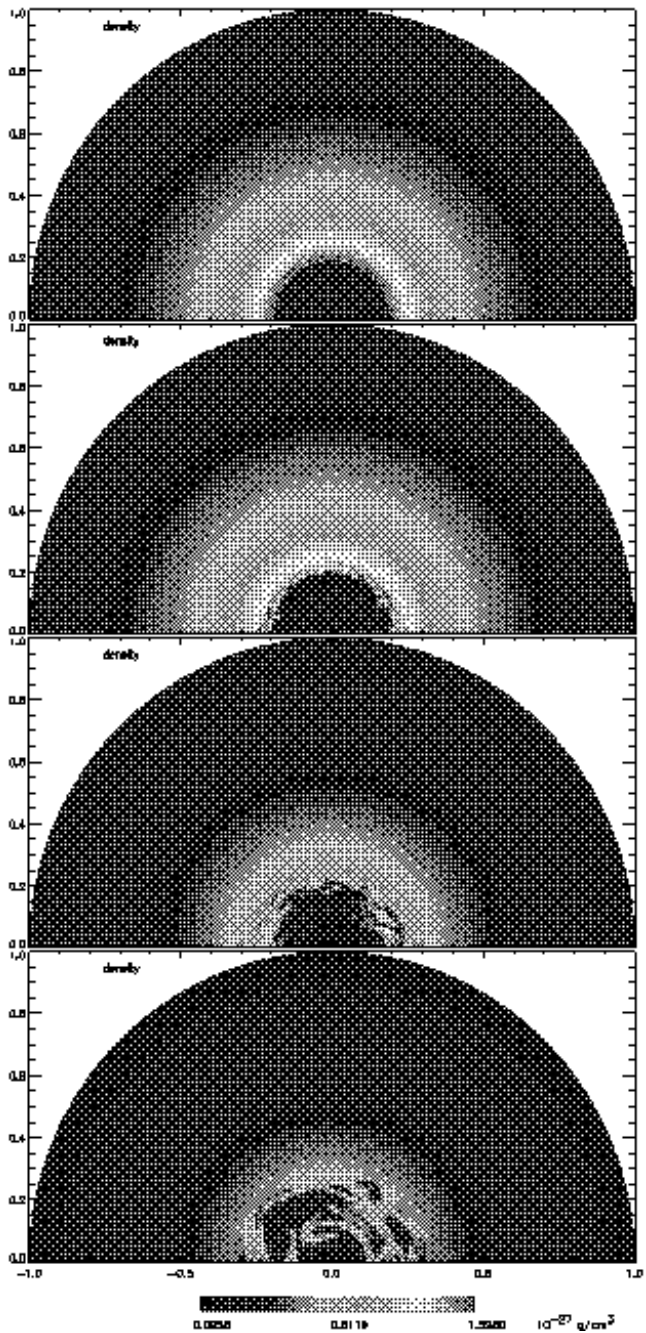}
\caption{Density maps of a spherical bubble (Model A, see text) at
four time steps. From top to bottom: Intial configuration, after 1.86
Gyrs, after 2.23 Gyrs and after 2.60 Gyrs. In this and all other
figures, except Fig. \ref{fig:modD}, unity on axes corresponds to 1
Mpc.}
\label{fig:modA}
\end{figure}

Fig. \ref{fig:modA} shows the results of our simulation for Model A:
a spherical bubble in the centre of the external density profile. We
did not perturb the bubble surface but only destroyed the symmetry by
raising the density of a single cell by a few percent. Thus the
configuration is very symmetric and stable. The high degree of
symmetry of the initial configuration prevents any large-scale
motions. Rayleigh-Taylor instabilities only start growing after about
1.8 Gyrs. The instability begins on small scales with the occurrence
of small `mushrooms' which develop at the density interface. These
mushrooms are characteristic for the Rayleigh-Taylor instability and
are also seen in laboratory experiments.  The small instabilities grow
and eventually lead to macroscopic mixing of the bubble material with
the ambient medium after another 400 Myrs. The growth of the
instabilities is consistent with theoretical predictions: Neglecting
the effects of compressibility, the growth rate, $\Gamma$, for an
instability with a characteristic wavelength $\lambda$ is
approximately $\Gamma^2 \sim g (\rho_{\rm a}-\rho_{\rm b})/\lambda
(\rho_{\rm a}+\rho_{\rm b})$, where $g$ is the gravitational
acceleration, and $\rho_{\rm a}$, $\rho_{\rm b}$ are the mass
densities of the ambient and the bubble gas densities, respectively
(e.g. Chandrasekhar 1961). \\

\begin{figure}
\plotone{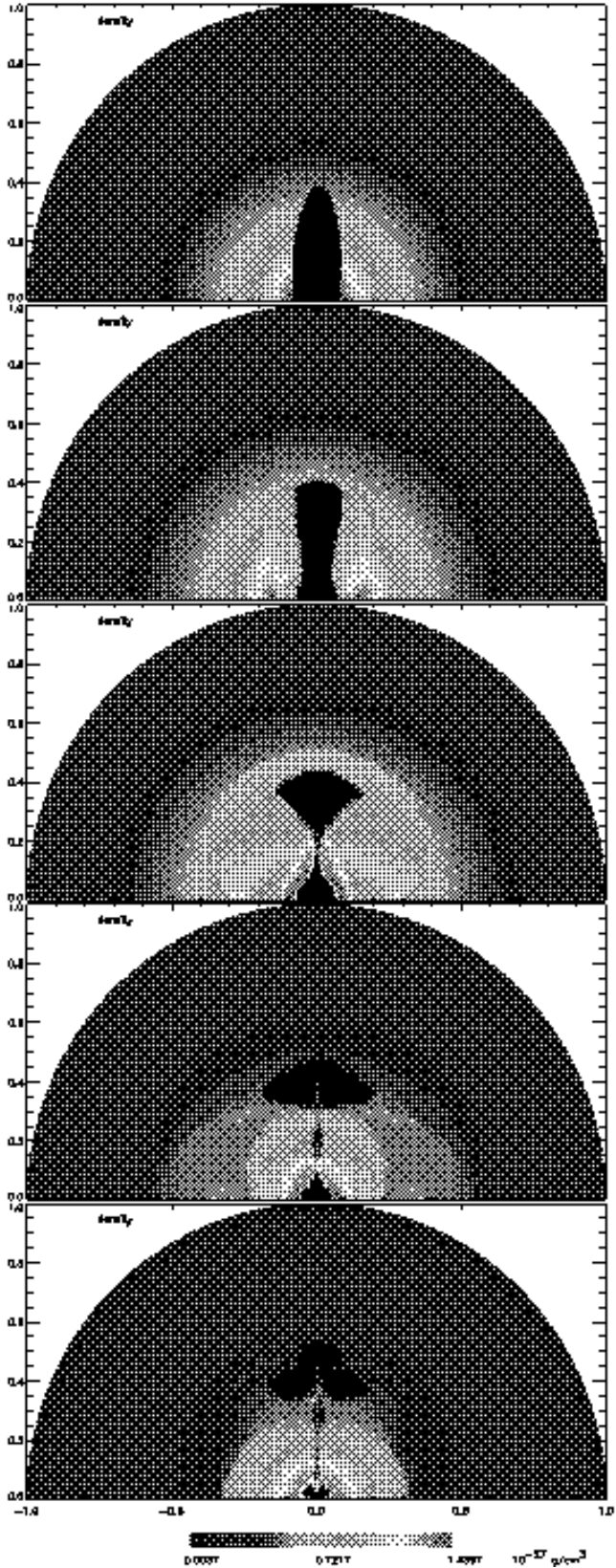}
\caption{Density maps of an elliptical bubble (Model B, see text) at
five time steps. From top to bottom: Initial configuration, after 46.5
Myrs, after 108.5 Myrs, after 217 Myrs and after 310 Myrs.}
\label{fig:modB}
\end{figure}

In Fig. \ref{fig:modB} we show the results for Model B. Due to its
elliptical shape this model is less symmetric than the spherical
bubble of model A. This leads to a faster growth of surface
instabilities on large scales. The largest pertubation that pinches
off parts of the bubble is already well developed after about 45
Myrs. Small-scale instabilities are visible along the edge of the
bubble at this time but they are far less significant than the
pinch. After 100 Myrs the mushroom shape is fully formed, leaving some
bubble material behind at the centre of the gravitational
potential. The mushroom cap evolves further into a torus with some
denser ambient gas being entrained in the torus centre. Denser
material surrounding the `trunk' of the mushroom is lifted up in the
wake of the ascending bubble. Finally, after about 300 Myrs
instabilities along the outer surface of the primary torus lead to a
split of the torus.

Several sound waves propagate through the ambient gas. The most
notable of these is a compression wave which at early times envelops
the top part of the bubble. More sound waves are visible at other
stages of the simulation but they are rather weak compared to the
primary wave.\\

\begin{figure}
\plotone{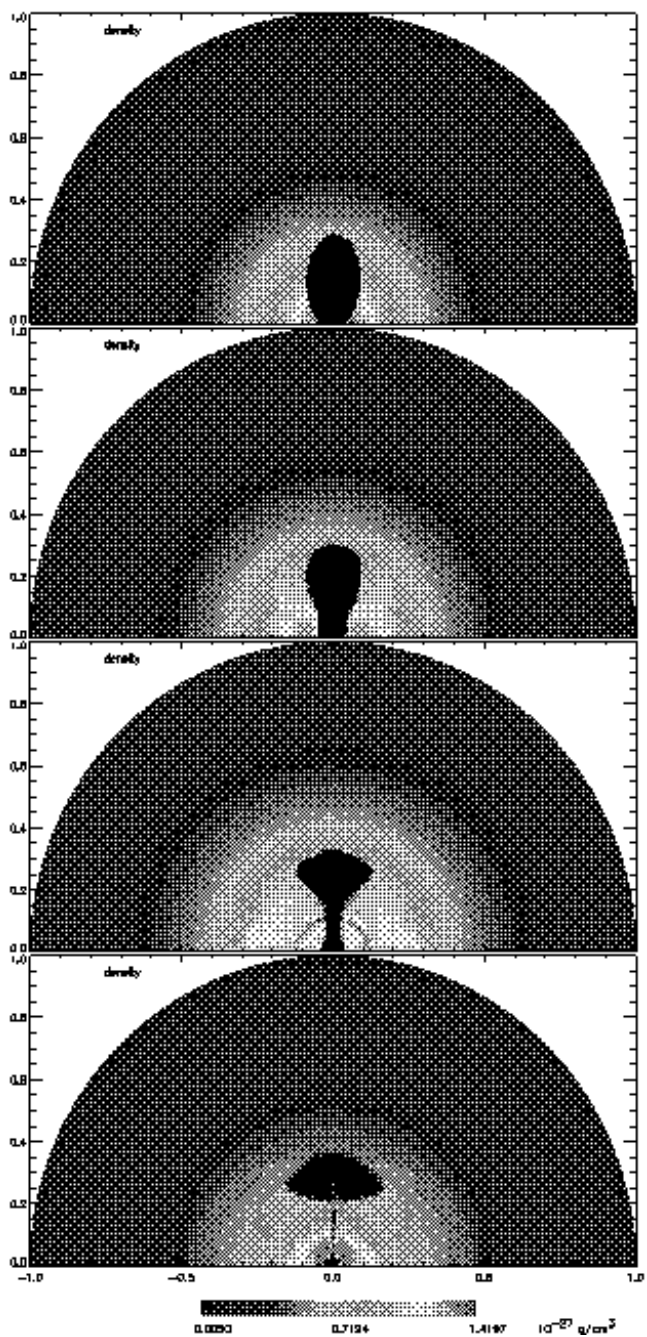}
\caption{Density maps of a `pre-pinched' elliptical bubble (Model C,
see text) at four time steps corresponding to the first four panels of
Figure \ref{fig:modB}. From top to bottom: Initial configuration,
after 46.5 Myrs, after 108.5 Myrs and after 217 Myrs.}
\label{fig:modC}
\end{figure}

In Sec. \ref{sec:simu} we discussed the possibility that a pinch at
the centre of the otherwise elliptical radio lobes may occur before
the start of the buoyant phase simulated here. In model C we simulated
the effect of such a pinch at the centre of the radio lobe (see top
panel of Fig. \ref{fig:modC}). The major axis of the bubble in model
C is shorter than that of model B. Nevertheless, as can be
seen by comparing Fig. \ref{fig:modB} and \ref{fig:modC}, the
evolution of the bubble is remarkably similar in the two models. The
existence of an already pinched shape does not significantly
accelerate the growth of the pinching instability. The bubble in model
C seems to pinch off at a somewhat lower position along the lobe than
that of model B. After the formation of the mushroom at around 200
Myrs, models B and C only differ in the sizes of their bubbles which
is mainly due to their different initial sizes.\\

\begin{figure}
\plotone{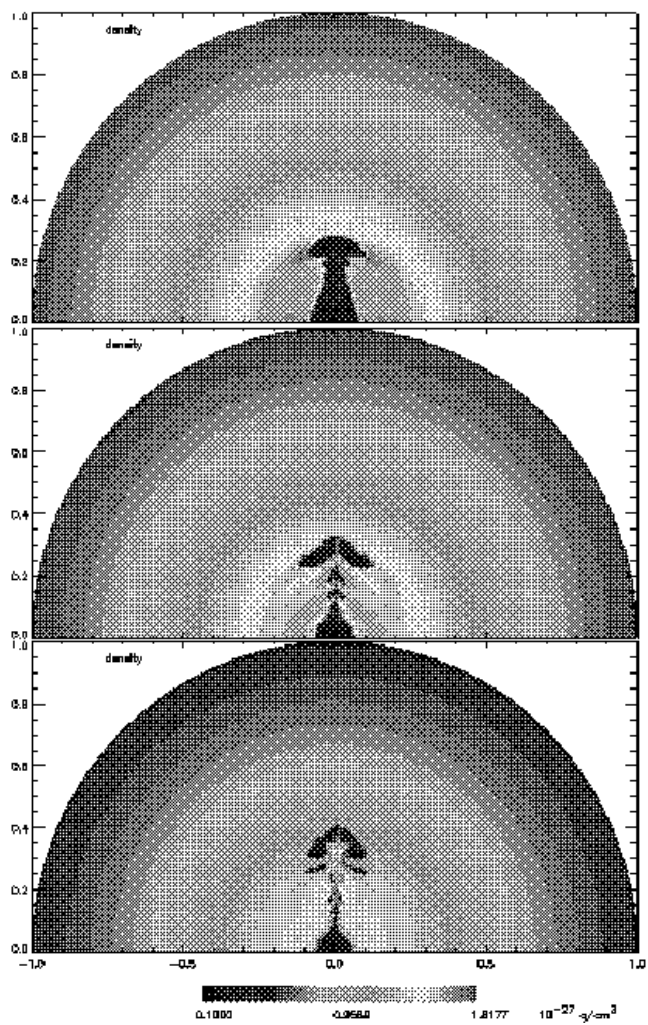}
\caption{Density maps of an elliptical bubble inside the core radius
(Model D, see text) at three time steps corresponding to the last
three panels of Fig. \ref{fig:modB}. From top to bottom: After 124
Myrs, after 217 Myrs and after 310 Myrs. In this figure unity on the
axes corresponds to 400 kpc.}
\label{fig:modD}
\end{figure}

All the radio lobes shown so far extend to twice the core radius of
the external density profile, $a_0$. The acceleration, $g$, due to the
underlying gravitational potantial rises linearly for distances small
compared to $a_0$ and falls off as $1/r$ at large distances $r$. The
acceleration reaches its maximum at $r=a_0$ and so all models
presented here, except for model D, simulate bubbles which initially
experience the maximum gravitational acceleration about halfway along
their lengths. Also, for most of their evolution the front ends of the
bubbles are immersed in a density profile roughly following a power
law distribution outside the core radius. In model D the bubble only
extends to half the core radius. Since the bubble is located inside
the core radius, the density profile of the surrounding medium does
not approximate a power law. Fig. \ref{fig:modD} shows the results
for this model. Note that this figure shows only the inner 400 kpc of
the cluster gas while all other figures extend to 1 Mpc. The initial
configuration is not shown as it is similar to a scaled version of the
first panel of Fig. \ref{fig:modB}. The bubble also develops a
mushroom shape but this happens slightly later than for the larger
bubbles. The pinching instability occurs further out and secondary
pinches develop along the rather thick trunk of the developing
mushroom. At the end of the simulation at least five separate tori are
discernible. The uplifted denser material is clearly visible in
the centre of the torus.\\

The simulations presented here extend the simulations performed by
Churazov et al. (2000b)\nocite{cbkbf00}. They employed a background
density profile that was derived from observations of the cluster
around Virgo A. Within this model they set up a spherical, buoyant
bubble which was off-set from the cluster centre. Again a rising torus
develops which splits into two tori at later times but no `trunk' is
observed. These results demonstrate how sensitive the detailed
morphology of the buoyant bubbles depends on the initial
conditions. Nevertheless, the development of a rising torus which
lifts up ambient material from the centre is generic.\\

Finally, we should briefly discuss the accuracy of these kinds of
finite-difference hydrodynamical simulations.  While the code can
simulate large-scale mixing due to Rayleigh-Taylor and
Kelvin-Helmholtz instabilities, it does not include real diffusion of
particles. Any observed diffusion is therefore entirely numerical. Any
sharp density discontinuities becomes less sharp as they are advected
across the grid due to discretization errors in the advection
scheme. For a test of the advection algorithm in the ZEUS code see
Stone \& Norman (1992).  Therefore, the small features in our
simulation are likely to be affected by these advection errors whereas
the larger features are not.

Second, numerical viscosity is also responsible for suppressing
small-scale instabilities at the interface between the buoyant bubble
and the surrounding gas.  To assess the effects of numerical
viscosity, we have repeated our simulations on grids with 100 $\times$
100 grid points. From our experiments we can conclude that ``global
parameters'' such as the position and size of the bubble as well as
the presence of ``toroidal'' structure are relatively insensitive to
the resolution. The detailed developments of the morphology on small
scales do depend on the resolution and the initial conditions.\\

\subsection{Rise velocity}

\begin{figure}
\plotthree{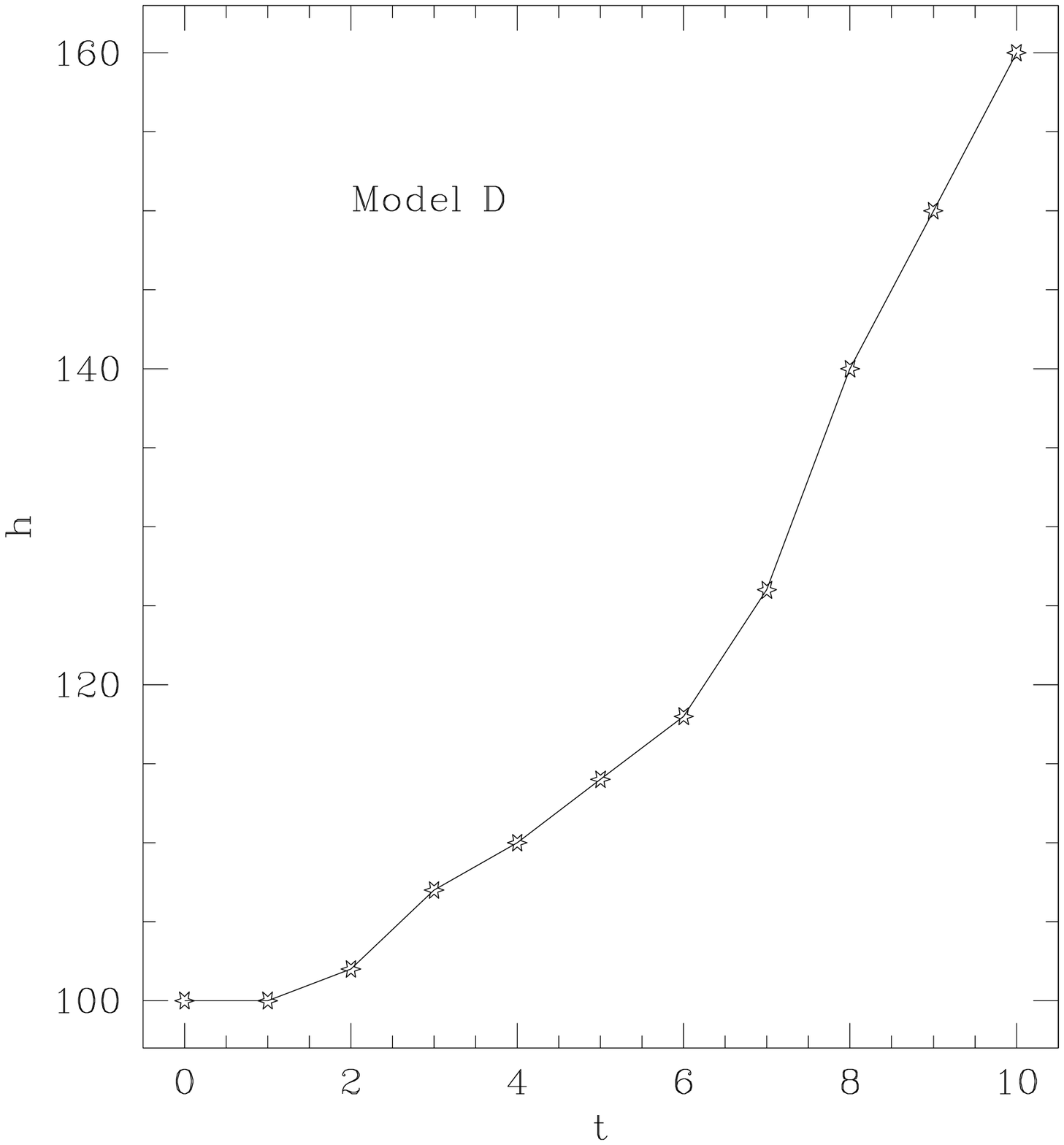}{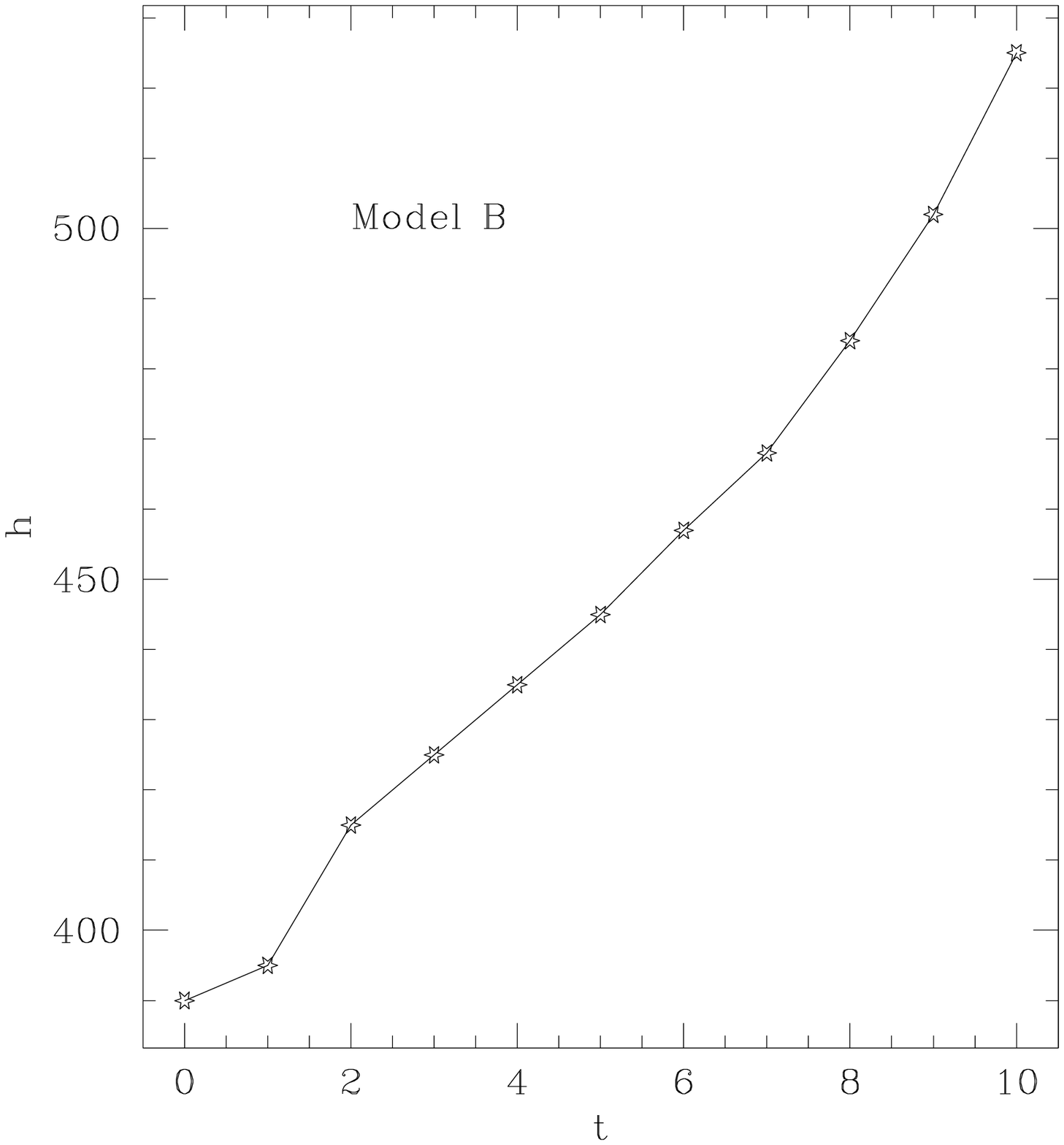}{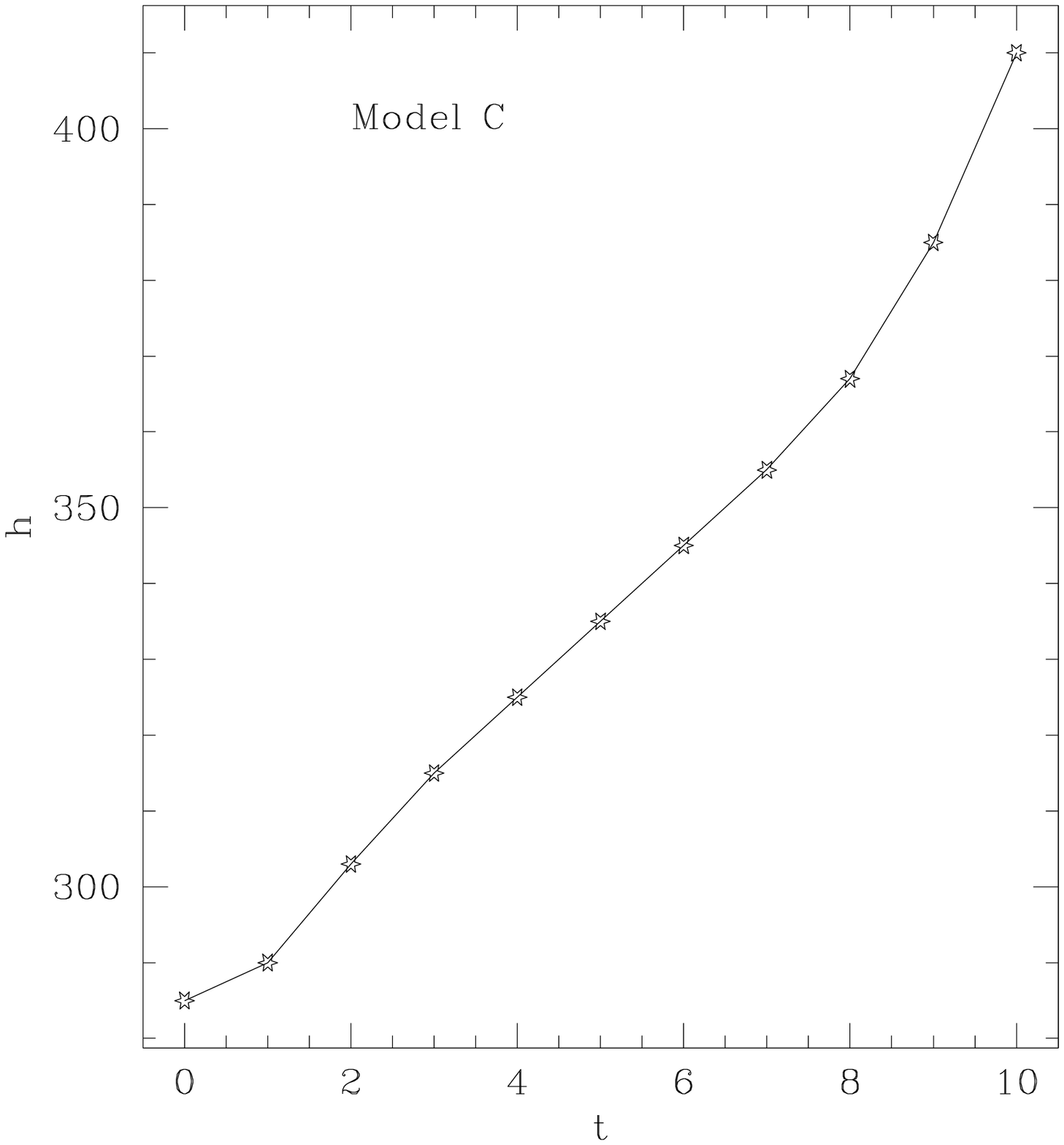}
\caption{Position of the bubble front as a function of time for models
B, C and D.}
\label{fig:rise}
\end{figure}

The simplest velocity estimate of the rising bubble can be
obtained by equating the ram pressure and buoyancy forces acting upon
a bubble (e.g. Gull and Northover 1973). The
buoyancy force is obviously:
\begin{eqnarray}
F_{\rm b}=Vg(\rho_{\rm a}-\rho_{\rm b}),
\end{eqnarray}
where $V$ is the bubble volume, $g$ is the gravitational acceleration,
$\rho_{\rm a}$ and $\rho_{\rm b}$ are the mass densities of the
ambient and the bubble gas densities respectively.  The drag force is
approximately given by
\begin{eqnarray}
F_{\rm d}\sim C \frac{1}{2} Sv^2\rho_{\rm a},
\end{eqnarray}
where $S$ is the cross section of the bubble. 
The numerical coefficient $C$ (drag coefficient) depends
on the shape of the bubble and the Reynolds number. 
For a solid sphere moving through an incompressible fluid this
coefficient is of the order 0.4--0.5 for Reynolds numbers
in the range $\sim 10^3$--$10^5$ (e.g. Landau and Lifshitz 1963). 
Thus the terminal velocity of the bubble is

\begin{eqnarray}
v\sim\sqrt{g\frac{V}{S}\frac{2}{C}\frac{\rho_{\rm a}-\rho_{\rm b}}{\rho_{\rm a}}}\sim\sqrt{g\frac{V}{S}\frac{2}{C}}.
\label{eqvr}
\end{eqnarray}

Here the factor of $(\rho_{\rm a}-\rho_{\rm b})/\rho_{\rm a}$ can be
dropped if the bubble density is low compared to the ambient gas
density. The expression for the terminal velocity can be further
rewritten using the Keplerian velocity at a given distance from the
cluster center: $v\sim \sqrt{(r/R)(8/3C)}v_{\rm K}$, where $r$ is the
bubble radius, $R$ is the distance from the center and $v_{\rm
K}=\sqrt{gR}$ is the Keplerian velocity. Of course the above formula
can be used only for order of magnitude estimates of the bubble
velocity. Fig. \ref{fig:rise} shows the position of the bubble front as a
function of time

\subsection{Ordered magnetic fields}

\begin{figure}
\plottwo{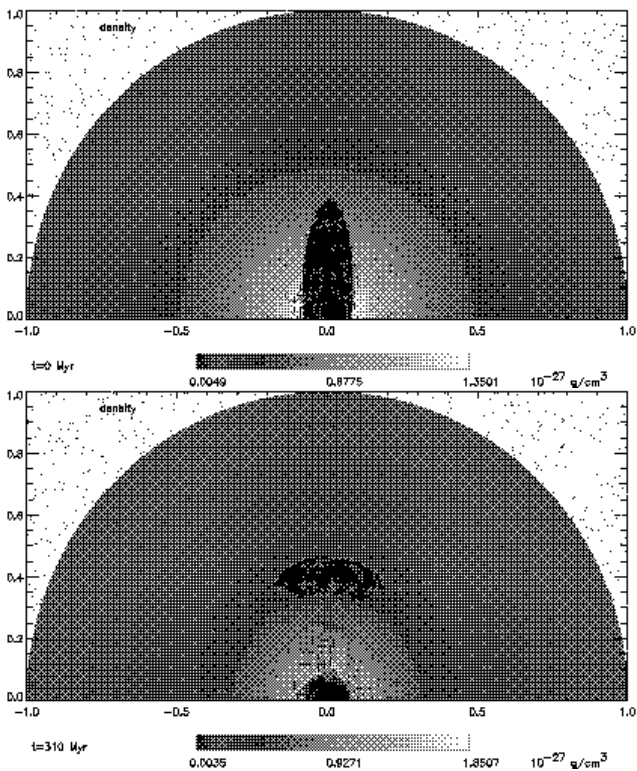}{fig5.epsi}
\caption{Density map of an elliptical bubble with a 'weak' magnetic
field aligned with the bubble surface (Model Bm1,
see text) at the the start of the simulation and after 310 Myrs. This
corresponds to the last panel of Fig. \ref{fig:modB}. The arrows
indicate the strength and direction of the magnetic field.}
\label{fig:modBm1}
\end{figure}

Fig. \ref{fig:modBm1} shows the final result for model Bm1. This
model is identical to model B shown in Fig. \ref{fig:modB} except
that we added a magnetic field that is initially aligned with the
bubble surface. The energy density of the magnetic field corresponds
to about 10\% of that of the thermal gas. The evolution of the bubble
is very similar to that in model B. The only slight difference is that
a larger bubble is left behind at the centre of the gravitational
potential in model Bm1. The magnetic field is not strong enough to
prevent the pinching instability and the bubble forms a mushroom as
before. However, the field is able to supress the break-up of the
rising torus during the late stages of the simulation. The magnetic
field is still aligned with the surface of the torus (see
Fig. \ref{fig:modBm1}) and stabilises it against Rayleigh-Taylor and
Kelvin-Helmholtz instabilities.

\begin{figure}
\plotone{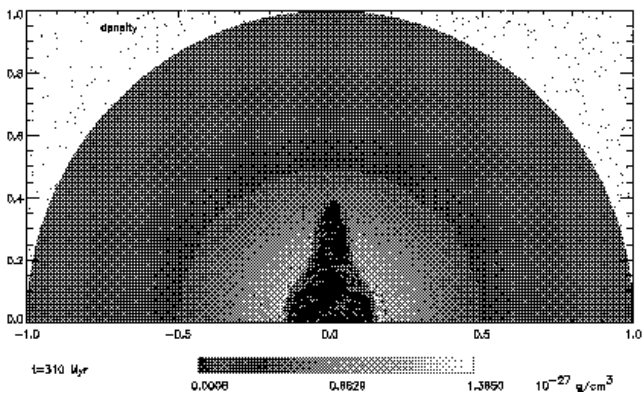}
\caption{Density map of an elliptical bubble with a strong magnetic
field aligned with the bubble surface (Model Bm2,
see text) at 310 Myrs after the start of the simulation. This
corresponds to the last panel of Fig. \ref{fig:modB}.}
\label{fig:modBm2}
\end{figure}

If the magnetic field, initially aligned with the bubble surface, is
stronger, it can provide enough surface tension to prevent the
pinching instability. Model Bm2 is identical to Bm1 but the energy
density of the magnetic field is by about a factor 3 higher. Fig.
\ref{fig:modBm2} shows the final stage of this simulation. The
magnetic field prevents the development of a pinch and the
configuration shown is essentially stationary. The circular shape of
the lower part of the bubble results from an expansion in this region
driven by the additional pressure exerted by the magnetic field. This
configuration resembles the spherical bubble of Model A but in model
Bm2 the magnetic field surpresses the small scale instabilities on the
surface. We conclude that the magnetic field can stabilise the bubble
against Rayleigh-Taylor instabilities. However, this requires a very
ordered field structure and a field strength which corresponds to a
significant fraction of the thermal pressure. It is unlikely that this
configuration is present in real radio lobes.

\section{Detection of buoyant bubbles}

It seems very likely that at the end of the lifetime of their jets
many powerful radio sources give rise to the buoyant structures in
clusters. Here we investigate whether they are directly observable.

\subsection{X-ray signature}

In our simulations we lowered the density of the bubble material by a
factor of 100 with respect to the density of the ambient gas. This
implies that in regions filled with bubble material the X-ray surface
emissivity is strongly reduced (see Churazov et
al. 2000b\nocite{cbkbf00}).  Zhidov (1977) showed that even in the
presence of surface tension small-scale surface instabilities may lead
to a fast break-up of the initial volume into smaller bubbles. Because
of the finite resolution of our simulations, we can not exclude that
microscopic mixing is efficient. However, the overall geometry of the
buoyant bubble, i.e. the development of a mushroom shape, is not
affected by this small-scale mixing. This implies that the decrement
in the X-ray emissivity within the buoyant bubbles may be much weaker
than is suggested by our simulations. Nevertheless, the dense material
uplifted by the rising torus from the centre of the external density
distribution will enhance the X-ray emissivity in the wake of the
torus (Churazov et al. 2000b\nocite{cbkbf00}). This may be identified
in some clusters with observed X-ray features extending radially from
the cluster centres (e.g., B\"ohringer 1995, Churazov et al. 2000b).

\subsection{Radio synchrotron emission}

\begin{figure*}
\plotonew{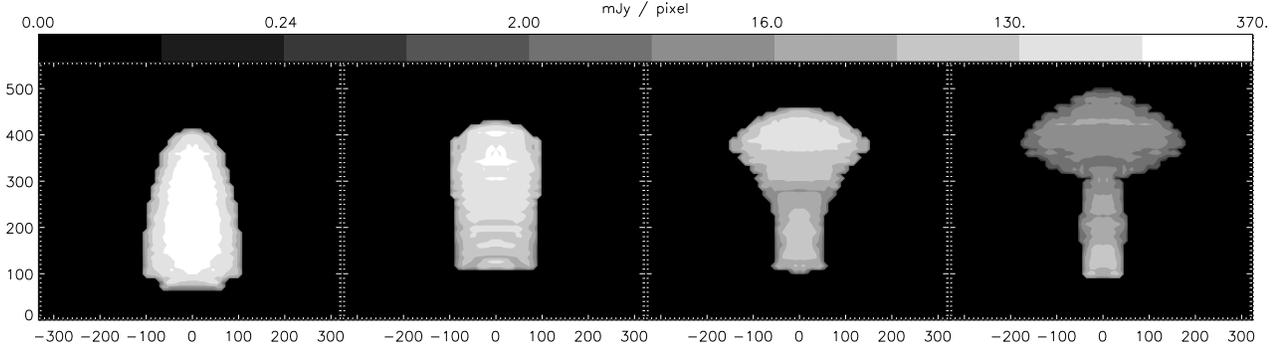}
\caption{Radio emission from model B assuming an initial magnetic
field strength of 2.3 $\mu$G and pressure equilibrium. The observing
frequency is 330 MHz and for the redshift we assume $z=0.1$. The
volume filling factor of the radio plasma is unity. The units
along the axes are kpc and the size of an individual pixel is
$9\times9$ kpc or roughly 3.7 arcsec$^2$ for the chosen redshift. From
left to right the panels show the bubbles at start of the simulation,
after 46.5 Myrs, after 108.5 Myrs and after 217 Myrs. They correspond
to the first four panels of Fig. \ref{fig:modB}.}
\label{fig:lowfield}
\end{figure*}

Using the assumptions summarised in Sec. \ref{sec:radsim}, we
calculated maps of the radio surface brightness of the buoyant bubbles
due to synchrotron radiation. In addition, we assumed an Einstein-de Sitter
universe with $H_{\rm o} =50$ km s$^{-1}$ Mpc$^{-1}$ and a redshift of
the simulated cluster of $z=0.1$. Except for the models that include a
large-scale magnetic field, Bm1 and Bm2, we are free to choose the
strength of the initial field. We considered two cases: A low field
with $B_{\rm initial} = B_{\rm CMB} / \sqrt{3} = 2.3 \mu$G, where
$B_{\rm CMB}$ is the equivalent magnetic field strength of the CMB,
and the equipartition case. In both cases the sum of the partial
pressures of the magnetic field and that of the relativistic particles
balances the local thermal pressure. The low magnetic field was chosen
because it allows for the maximum life time of the relativistic
particles (e.g. Churazov et al. 2000a\nocite{cfjb00}).\\

Fig. \ref{fig:lowfield} shows the result for the low field case and
an observing frequency of 330 MHz. In these maps the major axis of the
ellipse is in the plane of the sky. Even after $2\cdot 10^8$ years the
bubble is clearly detectable. The bubble appears to be broadened
compared to the density plots in Fig. \ref{fig:modB} because of the
lower numerical resolution provided by the advected tracer particles
compared to the computational grid. 

\begin{figure*}
\plotonew{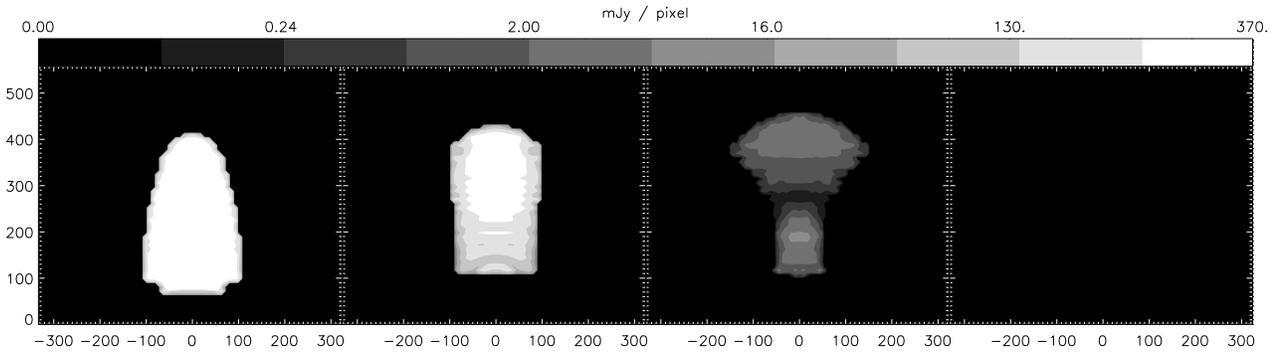}
\caption{Same as Fig. \ref{fig:lowfield} but assuming equipartition
between the magnetic field and the relativistic particles at the start
of the simulation.}
\label{fig:equipart}
\end{figure*}

Fig. \ref{fig:equipart} shows the same model as \ref{fig:lowfield}
but with the assumption that the magnetic field pressure and the
thermal pressure are in equipartition. In this case, the initial
magnetic field is stronger, $B\sim 15\mu$G. Consequently, in the early
phases the bubble is much brighter than in the low field case; the
peak surface brightness in the first panel of Fig.
\ref{fig:equipart} is above 1 Jy/pixel. This increased emission of
synchrotron radiation leads to faster energy losses of the
relativistic particles and so the bubble dims more rapidly than in the
low field case. After roughly $1.5 \cdot 10^8$ years it has faded
completely.\\

The radio surface brightness distributions resulting from the other
models considered in previous sections are qualitatively very similar
to that of model B. The total radio flux and the spectra depend more
on the assumptions for the strength of the magnetic field, the energy
distribution of the relativistic particles and the pressure in the
environment than on the morphology of the fluid flow. We also found
that the models which include large-scale magnetic fields showed
little difference in terms of total radio luminosity. Of course, in
these cases the magnetic field is very ordered and will give rise to
strong polarisation of the synchrotron emission. Rottmann et
al. (1996)\nocite{rmkw96} find polarised emission from the edges of
the buoyant large scale structure of M87. This seems to suggest some
ordering of the field in this object. However, this may simply reflect
the compression of the magnetic field along the leading edge of the
rising torus and is no indication of an initially ordered field.\\

In our calculations of the synchrotron radio emission we start with a
power-law energy distribution for the relativistic particles with a
rather high cut-off ($\gamma _{\rm max}=10^6$). This initial stage
corresponds to the point of pressure equilibrium between the radio
lobes and the ambient medium. If the relativistic particles in the
lobes of radio sources are only accelerated at the shock at the end of
an active jet then there will be no replenishment of these particles
once the jet is switched off. As pointed out above, at this stage the
lobes may still be overpressured with respect to the ambient
medium. The subsequent expansion phase without jet flow may well be
comparable or longer than the radiative lifetime of the more energetic
electrons. In this case, the high energy cut-off of the distribution
will be much lower than we assumed here. The lobes will then be
undetectable during their buoyant phase. 

Other effects such as {\em in situ} re-acceleration and diffusion of
particles between regions of low and high magnetic field may prolong
the radiative lifetime of the bubbles (for the case of M87 see Owen,
Eilek \& Kassim 2000\nocite{oek00}; Churazov et
al. 2000b\nocite{cbkbf00}). If these processes are significant, then
no conclusions on the source age can be drawn from the potentially
observed position of the break in the radio spectrum. A further
problem with these processes is that they require large amounts of
`hidden' energy in the relativistic particle population. Radio
emission in the GHz-range is emitted by electrons with Lorentz factors
of a few thousands. However, in re-acceleration processes usually many
more electrons at much lower energies are produced which dominate the
total energy density. They have much longer radiative lifetimes then
the `observable' electrons. Continued re-acceleration will therefore
lead to a build-up of pressure in the relativistic fluid, quickly
dispersing the emission region. In the case of an inhomogeneous
magnetic field in the lobes, the relativistic particles have to
diffuse slowly from regions of low magnetic field to high field
regions. If the typical diffusion time is short compared to the
radiative lifetime of the particles, then this kind of diffusion does
not significantly change the emission properties. In other words, if
diffusion is efficient, this case resembles that discussed above where
electrons are permanently in the high field region. For inefficient
diffusion only a small fraction of the relativistic particle
population is observable. The total energy density of the particles in
the low field regions, the `reservoir', is limited by the thermal
pressure of the surrounding fluid. This implies that, compared to the
radio maps shown in Figs. \ref{fig:lowfield} and \ref{fig:equipart},
such a source may radiate for longer but would be very faint.

Another possibility to extend the radiative lifetime of the buoyant
lobes is the local amplification of the magnetic field by shear flows
or other turbulent processes. These regions of stronger magnetic field
may then enable a comparatively unaged population of relativistic
particles to `light up'. However, even if we neglect adiabatic and
synchrotron energy losses altogether, a maximum lifetime of
relativistic particles is set by the IC scattering of the CMB. This
limits the time for which the radio plasma is observable to roughly
$2.3\cdot 10^9 (1+z)^{-4}$ years. The level of turbulence in the
bubble which amplifies the magnetic field is likely to decrease with
time. Also the total supply of relativistic electrons is still
depleted by consecutive shearing events. Therefore, the radio emission
again is expected to be faint at later times in this scenario.

\section{Conclusions}

Using hydrodynamical simulations, we investigated the evolution of
buoyant bubbles of different sizes and geometries in a galactic
gravitational potential. Common to all simulations with a
non-spherical bubble is the formation of a large mushroom-shaped lobe
that rises through the ambient medium. It is interesting to note that
the hot plasma remains confined within a single cloud for a considerable
time without requiring surface tension or magnetic fields. This is
also observed in the evolution of buoyant thermals which are formed,
for example, in strong nuclear explosions in the Earth's atmosphere.
In their wake the bubbles lift colder material upwards. \\
 
Differences between simulations with different initial geometries
become manifest in different rise velocities and somewhat different
morphologies of the buoyant bubble. We have selected a set of initial
conditions to show the ranges of rise velocities and morphologies that
can occur. The growth times as well as the morphologies of the
instabilities are quite different for spherical and elongated
bubbles. Only in the latter case a large `mushroom' forms. This
`mushroom' rises upwards as a whole before it succumbs to
instabilities on smaller scales. Spherical bubbles, however, evolve
very differently: Due to their symmetry, Rayleigh-Taylor instabilities
mix the material in the bubble with the ambient fluid {\it before} any
large-scale motions can occur. Our conclusions have to be diluted by
the usual caveats that come with finite-difference simulations. Our
simulations merely serve as toy models that demonstrate the underlying
physics and show the global morphology of the evolving bubbles. The
reader should bear in mind that the detailed morphologies on smaller
scales depend on the numerical procedure, the resolution of the grid
and the details of the initial model. Moreover, we only presented 2D
simulations which may produce artificial features. However, we have
repeated some of our simulations in 3D on a grid with much smaller
resolution and did not find significantly different results.\\

We also performed MHD simulations where the bubbles were filled with a
magnetic field that was parallel to the surface of the bubble. It was
found that the magnetic field helps to protect the rising bubbles
against secondary Rayleigh-Taylor and Kelvin-Helmholtz instabilities.
Only a strong and ordered magnetic field, where the magnetic
pressure is comparable to the thermal pressure, may suppress the
formation of the mushroom structure.\\

Finally, we calculated maps of the radio surface brightness due to
synchroton radiation. These maps were compiled for two different
assumptions about the magnetic field strength: (i) a low field, where the
initial field strength is roughly equal to the equivalent field
strength of the CMB, and (ii) a strong field, where the energy density
of the magnetic field and that of the relativistic particles are in
equipartition. In the weak-field case, the bubble is visible in the
radio for more than $2\cdot 10^8$ years, whereas in the strong-field
case the higher energy losses cause the radio image to fade after
$1.5\cdot 10^8$ years. We discuss various processes of re-acceleration
 and diffusion of relativistic particles as well as their influence on
the synchrotron radiation properties of the bubbles. We find that some
of these processes may prolong the phase of synchroton emission but
if so, the bubbles will be exceedingly faint at radio frequencies.

\section*{Acknowledgements}

It is a pleasure to thank Eugene Churazov for helpful discussions.

\end{document}